\documentclass[aps,prc,eqsecnum,twocolumn,amsmath,superscriptaddress]{revtex4}
\usepackage{graphics,graphicx,setspace,epsfig,color}
\usepackage[vcentermath]{}
\usepackage{epsf}
\usepackage{amscd}
\usepackage{amsmath,amssymb}
\usepackage{ulem}

\newcommand{\beq}{\begin{equation}}
\newcommand{\eeq}{\end{equation}}
\newcommand{\bea}{\begin{eqnarray}}
\newcommand{\eea}{\end{eqnarray}}
\newcommand{\nn}{\nonumber}

\newcommand{\hefour}{\mbox{${}^4$He\:}}
\newcommand{\htwo}{\mbox{${}^2$H\:}}
\newcommand{\hethree}{\mbox{${}^3$He\:}}
\newcommand{\hthree}{\mbox{${}^3$H\:}}
\newcommand{\liseven}{\mbox{${}^7$Li\:}}
\newcommand{\sing}{${}^1S_0$}
\newcommand{\trip}{${}^3S_1$}

\begin{document}
\preprint{INT-PUB-10-067}
\title[title]{Quark mass  variation constraints from Big Bang nucleosynthesis}
\author{ Paulo F.  Bedaque }
\affiliation{Maryland Center for Fundamental Physics \\
Department of Physics,
University of Maryland\\College Park, MD 20742, USA}
\author{ Thomas Luu }
\affiliation{N-Section, Lawrence-Livermore National Laboratory, Livermore, CA 94551, USA}
\author{Lucas Platter  }
\affiliation{Institute for Nuclear Theory, University of Washington, Seattle, WA 98195, USA}
\affiliation{Fundamental Physics, Chalmers University of Technology, SE-41296 G\"oteborg, Sweden}

%\date{}%
%\dedicatory{}%
%\commby{}%
% ----------------------------------------------------------------

\begin{abstract} 
  We study the impact on the primordial abundances of light elements
  created by a variation of the quark masses at the time of Big Bang
  nucleosynthesis (BBN). In order to navigate through the particle and
  nuclear physics required to connect quark masses to binding energies
  and reaction rates in a model-independent way, we use lattice QCD
  data and a hierarchy of effective field theories. We find that the
  measured $^4$He abundances put a bound of $ -1 \% \lesssim \delta
  m_q/m_q \lesssim 0.7\%$ on a possible variation of quark
  masses. The effect of quark mass variations on the deuterium
  abundances can be largely compensated by changes of the
  baryon-to-photon ratio $\eta$. Including bounds on the variation
  of $\eta$ coming from WMAP results and adding some additional assumptions
  further narrows the range of allowed values of $\delta m_q/m_q$.
 \end{abstract}

\maketitle

\section{Introduction}
In theories of physics beyond the standard model the standard
model parameters appear not as fundamental constants but as
derived quantities. In many of those theories the possibility then
arises that the values of the standard model ``constants" can vary over
time \cite{Marciano:1984rm}. It is then important to understand
which constraints the successes of standard cosmology -- which assumes
time independent constants -- poses on this purported time
variation. A natural place to look for a strong sensitivity to
a variation of fundamental constants is Big Bang nucleosynthesis (BBN)
since it satisfies two important criteria. First, BBN happened very
early in the universe's history, mostly when the universe was between
$3$ seconds and $3$ minutes old. Second, not only is standard BBN
understood at a few percent level but it is also very sensitive to
microscopic parameters such as nuclear binding energies and reaction rates
that are, themselves, very sensitive to certain standard model
constants. It is no surprise then that BBN has been used in the past
to study the variation of fundamental
constants~\cite{RevModPhys.75.403}. The purpose of the present paper
is to explore the BBN constraints on the variation of the masses of the
two lightest quarks, $m_u$ and $m_d$.

The binding of nucleons into light nuclei during BBN proceeds through
a number of reactions, some of which are in equilibrium with the
expansion of the universe and some that are not. After weak
reactions like $p+e^- \leftrightarrow n + \bar\nu$ are no longer in
equilibrium (i.e. {\it weak freezeout}), the ratio of neutron to
protons decreases due to neutron $\beta$-decay. If the formation of
light nuclei occurred in equilibrium, the most bound nuclei (among the
light ones this is \hefour) would form earlier and more
abundantly. The formation of \hefour can, however, only occur after
\htwo, \hethree and \hthree have been formed, since multinucleon fusion
reactions are essentially impossible at the relatively low densities
prevalent during BBN. Their number is small on the account that their
binding energies are small and it is not energetically favorable for them to
form until the temperature is low enough to be comparable to their
binding energies. Thus, the beginning of nucleosynthesis is delayed by
the shallowness of the deuteron binding energy, the so-called {\it deuterium
  bottleneck}. Since this shallowness is a product of delicate
cancellations between kinetic and potential energies, the binding of
the deuteron is an obvious place where a small change in quark masses
can significantly alter the primordial abundances. Notice that the
rate for the reaction $n+p\leftrightarrow d+\gamma$ is not small; it
is sufficient to keep the deuteron number in thermal/chemical
equilibrium. It is the equilibrium deuteron number that is too small
for them to collide and be assembled in larger nuclei. After the
deuteron number grows enough, the reactions leading to the formation
of \hefour proceed quickly and essentially all the neutrons present in
the beginning of BBN are assembled into \hefour nuclei. The timing
where this assembly starts (determined, among other things, by the
deuteron binding) is crucial as the neutron numbers are decreasing due
to neutron $\beta$-decay. Small amounts of \htwo, \hethree and \hthree
are left out of this process. Their numbers depend critically on
chemical non-equilibrium physics and the rates of the reactions,
including the initial $n+p\rightarrow d+\gamma$ reaction. Current
observation is not useful in measuring reliably the primordial
abundance of \hethree and \hthree. However, the abundance of \htwo,
and especially \hefour, are well measured and put a significant
constraint on any change of the standard BBN scenario.

A number of authors have previously considered the effect of quark
mass variations on the BBN predicted abundances
\cite{Kneller:2003xf,Dmitriev:2003qq,Coc:2006sx,
  Flambaum:2002wq,Dmitriev:2002kv,PhysRevC.76.054002,Berengut:2009js,
  Chamoun:2005xr,Landau:2004rj,Muller:2004gu,Flambaum:2002de,Calmet:2002yg,Dent:2001ga}. The main
difficulty to be surmounted is that the quark mass dependence of
binding energies, reaction cross sections and decay rates that are
input to BBN models are difficult to determine. For instance, modern
nuclear potentials can describe very well nucleon-nucleon phase
shifts. They can also be used to compute binding energies with enough
precision (with the help of phenomenologically motivated three-nucleon
forces fit to some observables) and cross section for few-nucleon
reactions. These potentials are, however, tuned to data obtained from
experiment where the quark mass has its current value. What is usually
done in estimating the effect of quark mass variation is to change the
parameters in these models where this dependence is easy to track. For
instance, the range of nuclear forces, given by $1/m_\pi$can be
changed through the relation $m_\pi^2 \sim m_q$. But the long distance
part of the potential, sensitive to this range, is actually a small
part of the nucleon-nucleon interaction. The medium and short range
parts also have a quark mass dependence and, while this dependence is
likely to be milder, its effect on the overall nucleon-nucleon
interactions is still large due to fine-tuned cancellations that are
responsible for, among other things, the shallowness of the
deuteron. In this paper we avoid, as much as possible, model-{\it
  dependent} approximations of the properties of nucleons and its
nuclear forces, relying solely on the symmetries of QCD and its
connection to nuclear physics through more general arguments. In
particular we use effective field theories (supplemented with lattice
QCD data) to connect the change in quark masses to the inputs used in
BBN simulations.

\bigskip
\begin{figure*}[tbp]
  \includegraphics[width=15cm]{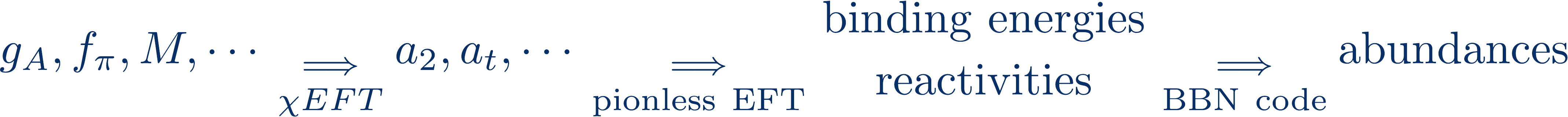}
%\vskip 0.15in
\noindent
\caption{The strategy used to determine quark mass dependence of BBN
  abundances.  At the far left, empirically determined LECs are used
  to constrain $\chi$EFT, allowing predictions of nuclear observables
  and determination of their quark mass dependence. This theory is in
  turn matched onto the pionless EFT, where subsequent calculations of
  binding energies and reactivities relevant to BBN are used as input
  into BBN codes.  }
\label{fig:bigpicture}
\end{figure*}  
%%%%%%%%%%%%%%%%%%

\subsection{Effective Field Theories}
At momentum scales $Q$ below $\Lambda_{QCD}\approx 1 $ GeV, the
relevant degrees of freedom in QCD are hadrons, not quarks and
gluons. Effective field theories (EFTs) for this momentum range (i.e. chiral
perturbation theory) were developed for the meson, one and
many-nucleon sectors. They are able to predict physical observables as
an expansion in the small parameters $Q/\Lambda_{QCD}$ and
$m_\pi/\Lambda_{QCD}$, Taking as inputs a few ``low energy constants"
(LECs), like pion decay constants and the nucleon mass in the chiral
($m_\pi=0$) limit.  These LECs, in turn, are determined from analyses
of experimental results. Effective field theories predict, for
example, the dependence of nucleon masses on the value of the quark
masses. This particular change, however, is very small and can be
neglected, except for its effect on the phase space for the neutron
decay and related weak process (see below). Lattice QCD calculations
reinforce the belief in a small quark mass dependence of nucleon
masses \cite{WalkerLoud:2008bp,WalkerLoud:2008pj}. On the other hand,
chiral perturbation theory for few-nucleon systems (referred to here
as $\chi$EFT) is in a less developed phase. First, there are
conceptual issues that preclude a reliable prediction of the quark
mass dependence of few-nucleon observables
\cite{Beane:2001bc}. Second, it has not been used extensively in
multi-nucleon systems and their reactions involving photons. To bypass
this difficulty we use a low-energy effective theory where all
particles, such as the pions, have been integrated out with, leaving
only the nucleon, photon, and neutrino degrees of freedom.  Known as
``pionless EFT", the momentum scales $Q$ relevant to this effective
theory are much smaller than the pion mass $m_\pi$.  This theory can
make non-trivial predictions because the states pertinent to BBN
(\htwo, \hethree, \hefour) are loosely bounded and the typical momenta
$Q$ of their constituents are significantly below $\sim m_\pi$ and
therefore within the regime of validity of the pionless EFT
\footnote{The shallowness of these bound states is related to the fine
  tuning in the s-wave two-nucleon scattering. In fact, the scattering
  length in the two spin channels \sing and \trip ($a_s\approx -22$ fm
  and $a_t\approx 5.4$ fm) are unnaturally large, much larger than the
  naive expectation $\approx 1/m_\pi = 1.4 $ fm.}. The pionless EFT is
very successful in predicting observables in the three-nucleon sector
and there is indication that the same is true in the four-nucleon
sector \cite{Platter:2004zs,Kirscher:2009aj}. Since the
$\alpha$-particle is the most bound of s-wave nuclei, its successful
description in the pionless EFT might indicate that the theory can be
useful in studying larger nuclei. Since the pionless EFT makes no use
of the QCD chiral symmetry, it cannot directly predict the quark mass
dependence of observables. The parameters of the pionless EFT, at the
lowest orders in the low energy expansion, are the threshold
nucleon-nucleon scattering parameters (e.g. scattering lengths,
effective ranges, etc\ldots). These few parameters have been studied
using $\chi$EFT and we can use them to predict their variation with
quark masses. In addition, some lattice QCD results confirm and
reinforce the $\chi$EFT predictions for scattering length dependences
on quark masses. We use these $\chi$EFT results as input parameters
for the pionless EFT. This allows us to obtain estimates for the quark
mass dependence of nuclear properties relevant to BBN. We will then
use this information in combination with a standard BBN code to
compute the light elements abundances in order to constrain the values
of the quark masses during the Universe's first minutes. Our strategy
of combining these two types of effective theories is summarized in
Fig.~\ref{fig:bigpicture}. We will now describe the stages of our
calculation.

\subsection{Scattering length dependence on quark masses}
Different versions of $\chi$EFT have been used by different authors to
study the quark mass variation of the nucleon-nucleon $S$-wave
scattering lengths. The results depend on the spin channel. In the
spin singlet \sing channel and at leading order (LO) on the
$m_\pi/\Lambda_{QCD}$ expansion, the calculation of the quark mass
dependence of the scattering length in the version of $\chi$EFT used
in \cite{Beane:2002xf} requires as inputs the chiral limit values of
the axial charge of the nucleon $g_A$, the decay constant of the pion
$f_\pi$, the nucleon mass $M$, the pion mass $m_\pi$ and the
coefficient of a two-nucleon contact term $C^0_s$ fitted to the
physical scattering length. Only the value of these quantities at the
physical value of quark masses is precisely known, but the difference
between them and their chiral limit values is a higher order effect
that can be neglected in a next-to-leading order (NLO) calculation. At
NLO a new constant $D^2_s$ appears (which is the coefficient of a
two-nucleon operator with no derivatives but one quark mass insertion)
as well as other constants contributing to the quark mass dependence
of $f_\pi, g_A$ and $M$. The value of $D^2_s$ is difficult to
disentangle from $C^0_s$ as both contribute equally to nucleon-nucleon
scattering at the physical value of the quark masses. They give,
however, different extrapolations to other values of quark
masses. They can be disentangled only through a study of processes like
deuteron-pion scattering or by the use of lattice QCD data (see
below). The strategy used in dealing with the lack of knowledge of the
value of $D^2_s$ is to estimate it using nai\"ve dimensional analysis
arguments. In Ref.~\cite{Beane:2002xf} $D^2_S$ was constrained
by requiring its absolute value not to be too much larger than $|C^0_S|$ while in
Ref.~\cite{Epelbaum:2002gb} $D^2_S$ itself was required to be of {\it
  natural} size.
%{\bf \large In contrast, in \cite{Epelbaum:2002gb} the
%  quark mass variation was determined \ldots}.
Fortunately, the
difference in the power counting schemes used in \cite{Beane:2002xf}
and \cite{Epelbaum:2002gb} has little impact on the dependence of the
scattering length on the quark masses
%{\bf \large of what?}
and the discrepancy between them can be
explained by the different assumptions about the reasonable range of
values for $D^2_s$. We will use the calculation described in
\cite{Epelbaum:2002gb} as those authors computed the quark-mass
dependence of both the deuteron binding energy and nucleon-nucleon
scattering lengths, since the deuteron binding energy is one of the
most important ingredients in the BBN calculation.

For a small variation of the quark mass we can read off figure 11
in Ref.~\cite{Epelbaum:2002gb} the slope (we use the more conservative
estimate where the change of the axial constant $g_A$ with quark
masses, parametrized by $\bar d_{16}$ is included):

\begin{equation}
\label{eq:meissnerB_2} \frac{m_q}{B_2} \frac{\delta B_2}{\delta
  m_q} = \frac{m_\pi}{2 B_2} \frac{\delta B_2}{\delta m_\pi} =
\frac{m_\pi}{2 B_2} \ (-0.085 \pm 0.027),
\end{equation}
where $m_q$ is the
average mass of the up and down quarks and we made use of the relation
$m_\pi^2 \sim m_q$.  Similarly, we use figure 12 in
\cite{Epelbaum:2002gb} to extract the variation of the spin singlet
\sing channel scattering length to find

\begin{equation}  \label{eq:meissnera_s}
\frac{\delta a_s}{\delta m_\pi} =
\frac{2m_q}{m_\pi}\frac{\delta a_s}{\delta m_q} =( -1.4 \pm 1.4
)\frac{\rm fm}{\rm MeV}. 
\end{equation}
Notice that a vanishing $a_s$ variation is consistent with these
extrapolations, a feature also seen in the extrapolation in
\cite{Beane:2002xf}. If $a_s$ were the only parameter determining the
change of abundances due to varying quark masses, BBN would impose no
constraint on possible quark mass variations.

Fully dynamical lattice QCD calculations of the nucleon-nucleon scattering
lengths have appeared in the last few years. They are still performed
at higher values of quark masses, too high for the effective theory
approach to be valid, so they are of limited value for our purposes.
Despite that, an attempt was made in \cite{Beane:2006mx} to use
$\chi$EFT to find the quark mass dependence of scattering lengths by
interpolating the lowest pion mass lattice data and the known
experimental value of the scattering lengths at the physical point.
At this point in time, the deuteron binding energy has not been
measured from lattice QCD. However, it is related, at leading order in
the effective theory, to the triplet scattering length that is
measured. Using the extrapolation in \cite{Beane:2006mx} and the
leading order relation $B_2=1/(M a_t^2)$ we find

\begin{equation}
\frac{m_q}{B_2}\frac{dB_2}{dm_\pi} = -0.14\pm0.13,
\end{equation}
in agreement with Eq.~(\ref{eq:meissnerB_2}). In the extrapolation
done in \cite{Beane:2006mx} another branch of allowed values of
$dB_2/dm_\pi$ appears. This additional band is excluded from the
purely EFT extrapolations in \cite{Beane:2002xf} and
\cite{Epelbaum:2002gb} and will be disregarded in this paper.

The allowed values for the $a_s$ quark mass dependence extracted from the extrapolation in \cite{Beane:2006mx} , namely
\begin{equation}
\frac{d a_s}{dm_\pi} = (-0.75\pm 1.0) \frac{\rm fm}{\rm MeV}
\end{equation}
are consistent with the ones above but are too loose to add any relevant constraint.

The remaining inputs of the pionless EFT, like three-nucleon
interaction parameters, effective ranges, nucleon magnetic moments,
etc\ldots, are not fine-tuned and therefore vary much less drastically
with the quark masses. Their contribution to the overall fusion cross
sections is also suppressed compared to $B_2$ and $a_s$. In the
present paper we will take them to be independent of the quark masses.

\subsection{Binding energies, reactivities and lifetimes}

We have used the pionless EFT to estimate the quark mass variation of
four quantities: the binding energies of the deuteron, \hthree,
\hethree, \hefour and the reactivity of the process $n+p\rightarrow
d+\gamma$. Similar calculations were carried out for \hthree in the
context of infrared limit cycles in
Refs.~\cite{Braaten:2003eu,Epelbaum:2006gd,Hammer:2007cs}. The
binding energies of larger nuclei, like \liseven, are important only
for the abundances for these larger nuclei. As it is not presently
possible to have a reliable estimate on the quark mass variation of
these binding energies, we keep them fixed and concentrate on the
abundances for the lighter nuclei \htwo and \hefour, confident that
they will not be significantly affected by the binding of $A>4$
nuclei. We also only include the variation of the reactivity of  proton-neutron capture  as this is the reaction that initiates BBN and is more
likely to have an impact on abundances (but, as we will see below,
this impact is minimal).  The binding energy of the deuteron is given
by Eq.~(\ref{eq:meissnerB_2}).

The calculation of three-nucleon and four-nucleon properties in the
pionless EFT requires as inputs the singlet and triplet scattering
lengths as well as one three-body observable, usually taken to be the
triton binding energy. This binding energy can be traded by the value
of a three-body force counterterm. The three-body force is also not
fine tuned and will therefore show only a weak dependence on the quark
masses that we will consequently neglect. Changing the two-body input
while keeping the three-body counterterm fixed provides then the
scattering length dependence of the three-nucleon system.  In other
words, the binding energies of the \hethree, \hthree and \hefour
nuclei are estimated by \beq\label{eq:Bnuclei}
\frac{m_q}{B_i}\frac{dB_i}{dm_q} =\frac{m_q}{ B_i} \left(
  \frac{da_s}{dm_q}\frac{dB_i}{da_s} +
  \frac{dB_2}{dm_q}\frac{dB_i}{dB_2} \right) \eeq where $B_i$ stands
for the binding energy of one of \hethree, \hthree or \hefour. The
values of the derivatives appearing in Eq.~(\ref{eq:Bnuclei}) were
computed using the pionless EFT:
\begin{eqnarray}
\frac{a_s}{B_3}
\frac{dB_3}{da_s} &=& 0.12,\ \ \ \ \
\frac{B_2}{B_3} \frac{dB_4}{dB_3} = 1.41,\nn\\
\frac{a_s}{B_4} \frac{dB_4}{da_s} &=& 0.037, \ \ \ \ \
\frac{B_2}{B_4} \frac{dB_4}{dB_2} = 0.74,\\
\end{eqnarray}
where $B_4$ is the \hefour binding and $B_3$ the \hthree or
\hethree binding energy. The weak dependence on $a_s$ is easily
understood when one notices that the typical momenta in these bound
states is of order $\sqrt{M B_i}$, which is much larger than
$1/a_s$. The dependence of $B_i$ on $a_s$ is a function of the
dimensionless parameter $\sim \sqrt{M B_i} a_s \ll 1$, and therefore
take to be zero.
% and, on a first approximation, we can take this parameter to vanish.

In order to account for the theoretical uncertainty in the EFT
calculation we assign an additional $10\%$ random variation to the
bindings of \hethree\ and \hthree\ (computed at NLO in EFT) and a
$30\%$ variation on the value of the \hefour\ binding (computed at LO
only), as will be shown more explicitly below.

The reaction $n+p\leftrightarrow d+\gamma$ was extensively analyzed in
Ref.~\cite{Rupak:1999rk} using a N$^4$LO calculation in the pionless
EFT. The inputs at this order are the scattering length $a_s$, the
deuteron binding energy, the corresponding effective range parameters,
the magnetic moments of the deuteron, and a single
two-nucleon-one-photon term fixed by experiment. We use the variation
of $B_2$ and $a_s$ given in Eqs.~\eqref{eq:meissnerB_2} and
\eqref{eq:meissnera_s} to compute, with the help of the explicit
formula in \cite{Rupak:1999rk}, the {\it relative} change in the
reactivity as a function of the temperature and use this as input for the BBN code. In
\cite{Dmitriev:2003qq} it was argued that the reactivity $\langle
\sigma v\rangle$ scales as $\sim B_2^{5/2} a_s^2$. We verified with
the explicit formula from \cite{Rupak:1999rk} that the scaling with
$B_2^{5/2}$ is indeed very well satisfied but that the scaling with $a_s^2$
does not work as well.

Finally, we discuss how quark mass changes affect the neutron lifetime
as well as the rates of other one-baryon weak reactions such as
$p+e^-\leftrightarrow n+\nu$. This effect arises from a modified
value of the axial charge $g_A$ and the neutron and proton masses,
which in turn dictate the allowed kinematic phase space for these weak
reactions. In fact, the neutron width is given by \cite{Fukugita:2003en}
\begin{equation}
 \Gamma = \frac{(G_F \cos\theta_c)^2}{2\pi^3}m_e^5 (1+3 g_A^2) f\left(\frac{\Delta}{m_e}\right),
 \end{equation}
 where $\Delta=m_n-m_p$ and $m_e$ are the mass splitting between
 neutron and proton and the electron mass, respectively, $g_A\approx
 1.26$ is the nucleon axial decay constant, $G_F$ the Fermi constant
 and $\theta_c$ the Cabibbo angle. The function $f(\Delta/{m_e})$ is
 \begin{equation}
f(w_0) = \int_1^{w_0} dw w \sqrt{w^2-1}(w_0-w)^2 \frac{2\pi
   \alpha}{\sqrt{w^2-1}}\frac{1}{1-e^{-\frac{2\pi
       \alpha}{\sqrt{w^2-1}}}}
\end{equation}
which describes the phase space and the Coulomb repulsion. The
variation of $\Gamma$ with the quark masses is given then by
\begin{equation}
\frac{m_q}{\Gamma} \frac{d\Gamma}{dm_q}
 = \frac{m_q}{f\left(\frac{\Delta}{m_e}\right)}
 \frac{d}{dm_q}f\left(\frac{\Delta}{m_e}\right) + \frac{m_q}{ 1+3
   g_A^2} 3\frac{d(g_A^2) }{dm_q}.
\end{equation}
The dependence of $g_A$ with the quark mass is given, at NLO in chiral
perturbation theory, by \cite{Bernard:1995dp}
\begin{equation}
g_A = g_A^0 \left [ 1- \frac{9 g_A^2
     m_\pi^2}{32 \pi^2 F^2} \log(\frac{m_\pi}{\Lambda})+ \frac{
     (g_A^2-4) m_\pi^2}{32 \pi^2 F^2} \log(\frac{m_\pi}{\Lambda'})
 \right],
\end{equation}
where $g_A^0$ is the chiral value of $g_A$, $F\approx 93$~MeV and
$\Lambda,\Lambda'$ are constants of order 1 GeV dependent on the
Gasser-Leutwyler coefficients \cite{Gasser:1983yg}. Numerically we find
\begin{equation}
\label{eq:dgAdmq} \frac{m_q}{1+3g_A^2} 3\frac{d (g_A^2)
 }{dm_q} = \frac{1}{2} \frac{m_\pi}{1+3g_A^2}3\frac{d(g_A^2) }{dm_\pi}
 \approx 0.2.
\end{equation}
 
The variation of the phase space $f(\Delta/m_e)$ with the quark masses
can be estimated as
 \begin{eqnarray}
 \frac{m_q}{f(\frac{\Delta}{m_e})}  \frac{f(\frac{\Delta}{m_e})}{dm_q} 
 &= &
 \frac{m_\pi}{2f(\frac{\Delta}{m_e})}\frac{df(\frac{\Delta}{m_e})}{dm_\pi}\\
 &=&
  \frac{m_\pi}{2f(w_0)}\frac{df(w_0)}{dw_0} |_{w_0
  =\frac{\Delta}{m_e}}\frac{d\Delta/m_e}{dm_\pi}\nn
\end{eqnarray}
 
The value of $f(w_0)$ and its derivative at $w_0=\Delta/m_e$ is found
numerically to be $1.64$ and $4.25$, respectively. The variation of
$\Delta/m_e$ with $m_q$ can be estimated by splitting $\Delta$ into a
strong interaction component $\Delta_s$ proportional to the up and
down quark mass difference (and, consequently, to the value of $m_q$)
and an electromagnetic piece $\Delta_{e.m.}$, largely independent of
$m_q$. Unfortunately, the electromagnetic part is due to short
distance effects and cannot be directly computed in a reliable
way. The best handle we have on its value comes from chiral
perturbation theory, where the up and down quark mass ratio, the meson
spectrum, and the best estimate of the nucleon $\sigma-$term are used
as inputs to extract $\Delta_s$. The value obtained for $\Delta_s$ in
this manner is consistent with that obtained from lattice QCD
calculation \cite{Beane:2006fk}. The difference between this value of
$\Delta_s$ and the measured value of the neutron-proton mass splitting
gives $\Delta_{e.m.} = -0.76\pm 0.30$ \cite{Gasser:1982ap}.
 
Chiral perturbation theory predicts a quark mass dependence of
$\Delta_s$ of the form $\Delta_s = A m_\pi^2 (m_d-m_u)/(m_d+m_u)$, a
formula valid up to NLO since the leading order loop contribution to
the nucleon mass cancels between the neutron and proton.  We then have
\begin{eqnarray}
\label{eq:dfdmq}
 \frac{m_q}{f(\frac{\Delta}{m_e})}  \frac{df(\frac{\Delta}{m_e})}{dm_q} 
 & =&\nn\\
&&\hspace{-2.3cm}=\frac{1}{f(w_0)}\frac{df(w_0)}{dw_0} |_{w_0=\frac{\Delta}{m_e}}  
  \frac{m_\pi}{2m_e} A \frac{m_d-m_u}{m_d+m_u} 2 m_\pi \nn\\
 &&\hspace{-2.3cm}=\frac{1}{f(w_0)}\frac{df(w_0)}{dw_0} |_{w_0=\frac{\Delta}{m_e}}  \frac{\Delta_s}{m_e}
  \approx 
  10.4 \pm 1.5.
 \end{eqnarray}
 Notice that we are taking both the up and down mass to vary while
 keeping the ratio $m_d/m_u$ fixed. As the dependence in
 Eq.~\eqref{eq:dfdmq} dominates over Eq.~\eqref{eq:dgAdmq}, we finally
 find
\begin{equation}
\label{eq:Gamma_variation} \frac{m_q}{\Gamma}
 \frac{d\Gamma}{dm_q} = 10.6\pm 1.5 .
\end{equation}
The quark mass variation of the neutron lifetime is relevant for our
calculation. In order to see that, let us remember that the neutron
number, after the weak interactions are decoupled, decreases until BBN
starts at $t\approx 168$ s. The suppression factor in standard BBN is
thus $ e^{-168/885}\approx 0.827$. A $5\%$ increase of quark masses
would lead, according to Eq.~(\ref{eq:dfdmq}), to a
  decrease,  of about $50\%$ in the neutron lifetime and the
suppression factor would change to $e^{-252/885}\approx 0.752$,
leading to a \hefour\ abundance change of about $10\%$ in the \hefour abundance, a variation comparable to the observational uncertainties.
  
 The rate of other weak reactions changes in a similar manner. The
 phase space integrals are more involved and are, in BBN codes,
 computed ``on the fly", taking the ratio $Q=\Delta/m_e$ as input. We
 calculated the variation of $Q$ as \bea\label{eq:weak_Q}
 \frac{m_q}{Q} \frac{dQ}{dm_q} &=& \frac{m_\pi}{2\Delta}
 \frac{d\Delta}{dm_\pi}
 =\frac{\Delta_s}{\Delta}\nn\\
 &\approx& 1.59\pm 0.23.  \eea

%%%%%%%%%%%%%%%%%%%%%%%%%%%%%%%%%%%%%%%%%%%%%%%%%%%%%%%%%%%%%%%%%%% 
\section{Results}
In order to deal with the highly non-linear dependence of the final
abundances on the quark masses and, at same time, to include estimates
of theoretical errors, we use a stochastic procedure. More
specifically, for a given quark mass variation $\delta m_q/m_q$, we
specify the binding energies of \htwo, \hthree, \hethree, \hefour, the
reactivity $\langle \sigma v\rangle$ for $n+p\leftrightarrow
d+\gamma$, the neutron lifetime $\tau$ and the phase space parameter
$Q$. All other BBN parameters are kept at their present values.

We have randomly generated a set of $300$ values of scattering lengths
$a_s$, deuteron bindings $B_2$ with a Gaussian distribution with mean
value and standard deviation given by
\begin{eqnarray}\label{eq:random_a}
\bar X & =& \left[ 1+ \frac{1}{2}\left(  \frac{m_q}{X}\frac{dX}{dm_q}|_+ + \frac{m_q}{X}\frac{dX}{dm_q}|_-\right) \frac{\Delta m_q}{m_q} \right] X^{\rm phys},\nn\\
\sigma_X & =&\left[ 1+ \frac{1}{2}\left(
    \frac{m_q}{X}\frac{dX}{dm_q}|_+ -
    \frac{m_q}{X}\frac{dX}{dm_q}|_-\right) \frac{\Delta m_q}{m_q}
\right] X^{\rm phys} ,\nn\\
\end{eqnarray}
where $X$ stands for either $a_s$ or $B_2$ and the ``+" and ``-"
subscripts refer to the higher and lower values of $dX/dm_q$ allowed
by Eqs.~\eqref{eq:meissnerB_2} and~\eqref{eq:meissnera_s}. The
variations of $a_s$ and $B_2$ are assumed to be uncorrelated. From the
ensemble of $a_s$ and $B_2$ obtained as above, we compute a
corresponding ensemble of binding energies using
Eq.~\eqref{eq:Bnuclei} and add to the result a $10\%$ (for \hthree\
and \hethree) or $30\%$ (for \hefour) relative random error in order
to take into account theoretical errors discussed in the previous
section. The binding energies of \hthree, \hethree\ and \hefour are
then given by
\begin{equation}
\label{eq:Bi}
\frac{B_i}{B^{\rm phys}}  = \left[ 1+ (1+t_i \xi_i)\left(\frac{a_s}{B_i} \frac{dB_i}{da_s} + \frac{B_2}{B_i} \frac{dB_i}{dB_2}\right)(a_s-a_s^{\rm phys}) )   \right],
\end{equation}
where $i$ indexes the three nuclei \hthree, \hethree and \hefour, the
superscript ``${\rm phys}$" stands for the present, experimental
values of the quantity, $\xi_i$ are Gaussian random variables with
central value $0$ and standard deviation equal to $1$, and $t_i$ is
the theoretical error of the extrapolation equal to $0.1$ (for \hthree
and \hethree) and $0.3$ (for \hefour).

Similarly, the reactivity $\langle \sigma v\rangle_T$ of the
$n+p\rightarrow d+\gamma$ reaction was computed as a function of the
temperature $T$ for the ensemble of $a_s, B_2$ values determined by
Eq.~\ref{eq:random_a} using the explicit expression for the cross
section from \cite{Rupak:1999rk}. The high order expansion of this
calculation in \cite{Rupak:1999rk} is accompanied with very small
theoretical errors that we subsequently neglect.

We also generated, for each value of $\delta m_q/m_q$, a set of 300
random values of $\tau$ and $Q$ whose distribution reflect the
discussion in the previous section. More specifically, these values
were generated through the formula
\begin{eqnarray}
\label{eq:tauQ}
\frac{1}{\tau} &=&\frac{1}{ \tau^{\rm phys}} \left[1+(10.6 + 1.5 \xi) \frac{\delta m_q}{m_q}\right],\nn\\
Q &=& Q^{\rm phys} \left [   1 +  (1.59 + 0.23 \xi)\frac{\delta m_q}{m_q}\right],
\end{eqnarray}
where $\xi$ is a Gaussian random variable with central value $0$ and
standard deviation $1$. Notice that this $\xi$ is independent of the
$\xi_i$ used in the determination of the binding energies but the same
$\xi$ is used in both $\tau$ and $Q$ since the leading theoretical
uncertainties on both quantities stem from the same determination of
the $\sigma$-term.

For a given value of $\delta m_q/m_q$, a set of values for $B_2, B_{
  ^3H}, B_{^3He}, B_4, \langle \sigma v\rangle_T$ was paired one of
set of $\tau$ and $Q$ values and used in a standard BBN code.  The BBN
code we have used in our analysis is based on
Refs.~\cite{Wagoner:1972jh,Kawano:1992ua} and is publicly
available~\cite{sakarcite}. The code was modified to accept
temperature-dependent variations in the reactivity corresponding to
the $n+p\rightarrow d+\gamma$ reaction and the rate of weak
interaction processes was changed according to
Eq.~(\ref{eq:Gamma_variation}) and Eq.~(\ref{eq:weak_Q}).  The
Q-values of all BBN reactions with \htwo, \hthree, \hethree, and
\hefour\ as either parent or daughter products of reactions were
allowed to vary in accordance with the changes in binding energies of
these nuclei.  The baryon-to-photon ratio $\eta$ was changed over a
range discussed below.  Otherwise, the standard input parameters were
used in our BBN simulations.

%%%%%%%%%%   He4vsH2plot _etas  %%%%%%%%%%%%%%%
\bigskip
\begin{figure}[tbp]
  \centerline{\includegraphics[width=7cm]{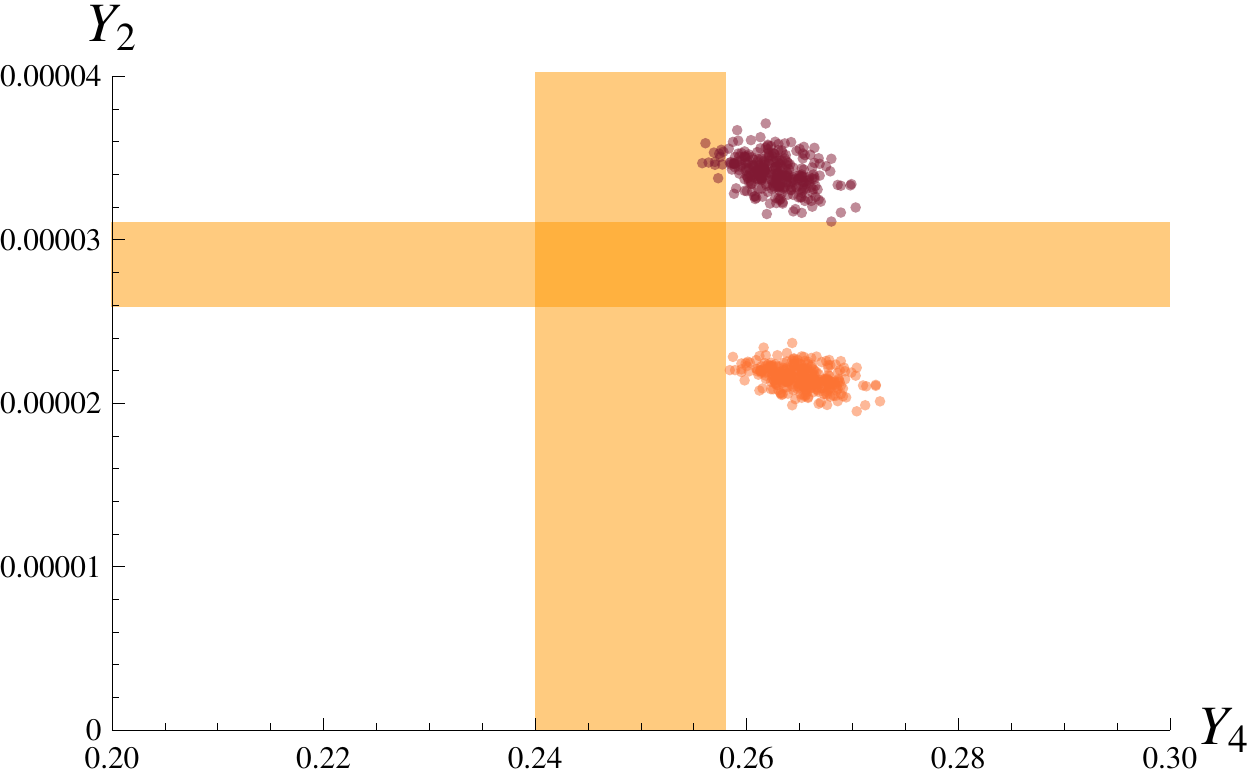}}
%\vskip 0.15in
\noindent
\caption{The yellow bands show the (1-$\sigma$ allowed abundances for
  \hefour and \htwo. The two clouds show the result of 300
  simulations, both with $\Delta m_q/m_q$ = -1 \% but two different
  values of $\eta_{10}$. The lower cloud (ochre online) is the result
  of taking $\eta_{10}=6.23$ and the upper cloud (burgundy online) the
  value $\eta_{10}=4.60$. There is very little change on the \hefour
  yield but the deuterium yield changes enough to render the deuterium
  abundance useless in putting a constraint on $\Delta m_q/m_q$.}
\label{fig:etas}
\end{figure}  
%%%%%%%%%%%%%%%%%%
The main feature seen in the simulations is that a variation in $\eta$
shifts the deuterium abundance but has little effect on the \hefour\
yields (see Fig.~(\ref{fig:etas})). A larger value of $\eta$ implies
in a larger baryon density, a more complete burning of the neutrons
into \hefour \ nuclei and a smaller deuterium abundance. As a
consequence, in the absence of a restriction on the value of $\eta$
from other considerations, the deuterium abundance does not put any
constraints on the range of allowed quark masses variations.

Additional constraints on the value of $\eta$ come from studies of the
large-scale structure of the Universe. The actual numerical value of
the constraints, however, depends on assumptions made in these
analyses, including assumptions on the initial spectrum of
fluctuations. For instance, the lower range of the determination of
$\eta_{10} = 4.79 \pm 0.019$ in \cite{Hunt:2007dn} and the central
value of the determination of $\eta_{10} = 6.23 \pm 0.17$ in
\cite{Dunkley:2008ie}, are shown for $\delta m_q/m_q$ in
Fig.~(\ref{fig:etas}). A similar plot results for other values of
$\delta m_q/m_q$. Consequently, any reasonable change in the deuterium
abundance can be accommodated by a change in the value of
$\eta_{10}$. If we restrict ourselves to the much narrower range
$\eta_{10} = 6.23 \pm 0.17$ \cite{Dunkley:2008ie}, the deuterium
abundances can play a role. However, the values in the range
$\eta_{10} = 6.23 \pm 0.17$ are in tension with the observed deuterium
abundances. BBN, by itself, prefers the slightly lower range $5.1 <
\eta_{10} < 6.5$, at the $95\%$ confidence level
\cite{Nakamura:2010zzi}. Thus, even with the current physical values
of $m_q$, the predicted deuterium abundance lies just outside the
$1-\sigma$ band, making it difficult to distinguish the allowed and
forbidden values of $m_q$ based on $Y_2$.  Therefore, to proceed
further, we disregard the deuterium abundances and look at how the
\hefour\ abundances change with the quark masses.

%%%%%%%%%%   He4vsH2plot   %%%%%%%%%%%%%%%
\bigskip
\begin{figure}[tbp]
  \centerline{\includegraphics[width=7cm]{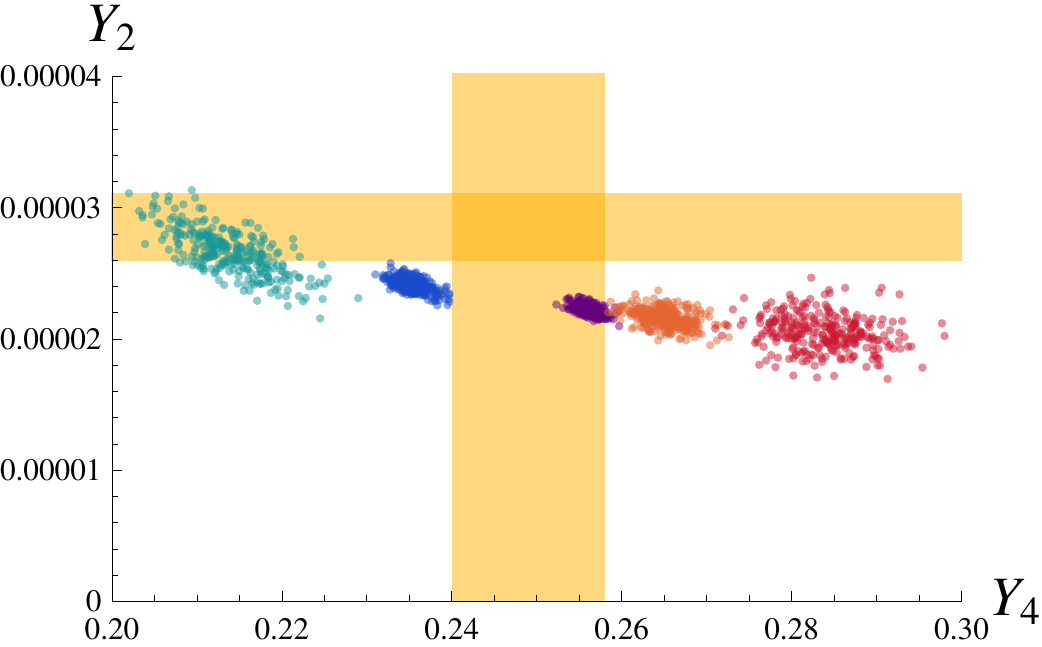}}
%\vskip 0.15in
\noindent
\caption{The yellow bands show the (1-$\sigma$ allowed abundances for
  \hefour and \htwo.  The five clouds show the result of 300
  simulations at each one of the values (from left to right): $\Delta
  m_q/m_q = 2\%$ (green online), $\Delta m_q/m_q = 0.7\%$ (blue
  online), $\Delta m_q/m_q = -0.5\%$ (purple online), $\Delta m_q/m_q
  = -1\%$ (ochre online), and $\Delta m_q/m_q = -2\%$ (red online).  }
\label{fig:He4vsH2plot}
\end{figure}  
%%%%%%%%%%%%%%%%%%
In Fig.~(\ref{fig:He4vsH2plot}) we show the result of changing the
quark masses by five values: $2\%, 0.7\%, -0.5\%, -1\%$ and $-2\%$,
all corresponding to $\eta_{10}=6.23$.  Each one of these values of
$\delta m_q/m_q$ is represented by a cloud of points in the $Y_4
\times Y_2$ plane. The spread between the 300 points in each cloud
accounts for the theoretical uncertainties in the extrapolation of the
parameter inputs as described by Eqs.~(\ref{eq:Bi}) and
(\ref{eq:tauQ}). The tendency is for a smaller $Y_4$ for larger values
of $m_q$. Two main mechanisms account for this general trend. First,
large values of $m_q$ imply larger values of $\Delta$, as well as a
larger phase space for neutron decay and therefore shorter neutron
lifetime. Consequently, more neutrons decay by the time BBN starts the
assembly of \hefour, resulting in smaller \hefour\ yields. In
addition, Eq.~(\ref{eq:meissnerB_2}) shows that a larger $m_q$ implies
a smaller $B_2$. The deuteron, being less bound, takes longer to form,
delaying the onset of \hefour\ formation and giving even more time for
the neutrons to decay, reducing further the \hefour\ yields.  There is
also a weak tendency to have smaller $Y_2$ for smaller $m_q$, a trend
not so easily explained.

Based on the data shown on Fig.~(\ref{fig:He4vsH2plot}) we put a bound
on the allowed values of quark mass changes at \beq -1 \% \lesssim
\frac{\Delta m_q}{m_q} \lesssim 0.7\%, \eeq which is the main result
of this paper. We refrain from assigning a numerical value to the
uncertainty in this estimate, as an attempt in this direction would
require us to assign a precise statistical meaning to our theoretical
uncertainties. While there are reasons to take these uncertainties
seriously at the qualitative level, we believe them to be superior to
the model calculations used previously.  
%%%%%%%%%%%%%%%%%%%%%%%%%%%%%%%%%%%%%%%%%%%%%%%%%%%%%%%%%%%%%%%%%%%%
\section{Conclusion}
We have estimated the abundances of \htwo and \hefour\ produced in the
standard BBN scenario under the assumption that the light quark masses
were shifted at the BBN time from their present values. In order to
perform this calculation we have used input from several effective field
theories as well as lattice QCD results to connect the quark mass
variation to the relevant nuclear physics pertinent to BBN. We found
that a variation beyond the $-1 \% \lesssim \frac{\Delta m_q}{m_q}
\lesssim 0.7\%$ range to be likely inconsistent with the observed
abundances.

Two of the BBN parameters played the largest roles in changing the
light element yields: the deuteron binding energy $B_2$ (with the
\hthree, \hethree and \hefour \ binding energies strongly correlated
with $B_2$) and the neutron lifetime. The dependence of the neutron
lifetime on the quark mass values is well constrained by theory. The
variation of the deuteron binding is, however, much less constrained
and several venues of further progress are clearly visible (for a very
recent study, see \cite{Chen:2010yt}). Lattice QCD calculations of
nucleon-nucleon interactions, even if performed at unphysical values
of $m_q$, would go a long way in narrowing these constraints. As long
as they are performed with quark masses low enough to be within the
region of validity of the chiral nuclear EFT, they determine reliably
the value of parameters of the EFT necessary for the extrapolation of
the deuteron binding energy. The binding energies of \hthree,
\hethree\ and, specially \hefour, can and should be computed in the
pionless nuclear EFT to higher orders so that theoretical
uncertainties associated with these quantities decrease. Finally, a
better understanding of the quark mass variation of other threshold
parameters like effective ranges, magnetic moments, etc\ldots , would
also allow for a more precisely constrained calculation of the binding
energies on nuclei larger then \hefour.

A number of other works have also considered the effects of a
variation of the quark masses on properties of light nuclei. For
example, in Ref. \cite{PhysRevC.76.054002} this effect was
implemented by a change of the pion mass in the phenomenological model
interaction employed in the calculations. Such an interaction is only
able to capture the true quark mass dependence to a limited degree
since it is not constructed as a systematic expansion in powers of
$m_\pi/\Lambda_{QCD}$. In particular, quark mass dependent
short-distance contact operators (such as the D-term) discussed in the
above text do not appear in standard phenomenological
interactions.

Since we are not presently able to obtain reliable values for the
${}^7Li$ binding energies, the ${}^7Li$ abundances we compute are not
very meaningful and were not used to put constraints on
the quark mass variations. Future advances in the nuclear pionless
effective theory may change this and allow us to address the ``Lithium
problem" as a signal of quark mass variation.

Finally, it should be pointed out that in models of physics beyond the
standard model, the value of the quark masses are derived quantities
and variations of them may well be correlated with other
quantities. In particular, it may seem unnatural to expect the masses
of different quark flavors to vary together, unless this variation is
being driven by a change in the Higgs vacuum expectation value. If
that is the case, a change in the quark masses will be correlated with
changes in the vector boson masses, changing the strength of strong
interaction at low energies. The effect of those changes on BBN can be
easily tracked in a manner similar to what was done in this paper. We
plan to consider BBN bounds on the Higgs vacuum expectation value
change in a future publication.

%%%%%%%%%
\acknowledgments

We thank T. Cohen for discussions. P.B. was supported by the
U.S. Department of Energy under Grant No. DE-FG02-93ER-40762. The work
of TL was performed under the auspices of the U.S.~Department
of Energy by Lawrence Livermore National Laboratory under
Contract DE-AC52-07NA27344 and the UNEDF SciDAC grant
DE-FC02-07ER41457. L.P. was supported by the Department of Energy under grant
number DE-FG02-00ER41132.

%%%%%%%%%%

%%%%%%%%%
%\bibliographystyle{h-physrev}
%\bibliographystyle{apsrev}
%\bibliography{BBN_variation} 

%%%%%%%%%%%%%%%%%%%%%%

\end{document}